\documentclass[doublecol,linenumbers]{epl2} 
\usepackage{graphicx}  
\usepackage{subfigure}
\usepackage{color}

\newcommand{\bq}{\begin{equation}}
\newcommand{\eq}{\end{equation}}
\newcommand{\ba}{\begin{eqnarray}}
\newcommand{\ea}{\end{eqnarray}}

\newcommand{\ul}{$\mu$l}
\newcommand{\um}{$\mu$m}

\title{Reduced adhesion between cells and substrate confers selective advantage in bacterial colonies}
\shorttitle{Reduced adhesion confers selective advantage}  
\author{Craig Watson\inst{1}, Paul Hush\inst{1}, Joshua Williams\inst{1}, Angela Dawson\inst{1}, Nikola Ojkic\inst{1}, Simon Titmuss\inst{1}, and Bartlomiej Waclaw\inst{1,2}}

\shortauthor{C. Watson \etal}

\institute{                    
	\inst{1} School of Physics and Astronomy, The University of Edinburgh, Peter Guthrie Tait Road, Edinburgh EH9 3FD, United Kingdom\\
	\inst{2} Centre for Synthetic and Systems Biology, Edinburgh EH9 3FD, United Kingdom
} 

\abstract{Microbial colonies cultured on agar Petri dishes have become a model system to study biological evolution in populations expanding in space. Processes such as clonal segregation and gene surfing have been shown to be affected by interactions between microbial cells and their environment. 
In this work we investigate the role of mechanical interactions such as cell-surface adhesion.
We compare two strains of the bacterium {\it E. coli}: a wild-type strain and a ``shaved'' strain that adheres less to agar. We show that the shaved strain has a selective advantage over the wild type: although both strains grow with the same rate in liquid media, the shaved strain produces colonies that expand faster on agar. This allows the shaved strain outgrow the wild type when both strains compete for space. We hypothesise that, in contrast to a more common scenario in which selective advantage results from increased growth rate, the higher fitness of the shaved strain is caused by reduced adhesion and friction with the agar surface. 
}


\begin{document}

\maketitle

\section{Introduction}


Micro-organisms -- a \um-sized single-celled organisms that usually replicate by binary fission -- are the most numerous organisms on Earth. They represent all domains of life (bacteria, archaea, and eukaryota) and take many forms (spherical, rod-shaped, spirals) \cite{schaechter_microbe._2006}. 

When provided with sufficient nutrients and a surface to attach to, a single microbial cell can create a macroscopic (mm- to cm-sized) colony in just one day. Microbial colonies growing on agar Petri dishes are a beautiful example of self-organisation \cite{ben2000cooperative}. Depending on the type (species) of microbe and growth conditions, colonies of different shapes are produced: compact, circular-symmetric colonies, concentric rings, ``curly-hair'' colonies, spirals, terraces, and branched, tree-like colonies \cite{shapiro_significances_1995,ben2000cooperative}. Experimental and theoretical work \cite{ben-jacob_generic_1994,kawasaki_modeling_1997,golding_studies_1999,xavier_social_2009,farrell_mechanically_2013,ali_scale-invariant_2013,giverso_branching_2015} has shown that these complex patterns often arise through a simple, reaction-diffusion mechanism due to the interplay between growth, nutrient consumption, death, chemotaxis, as well as other interactions between microbial cells.

\begin{figure}
	\includegraphics[width=\columnwidth, trim=0 200 0 0]{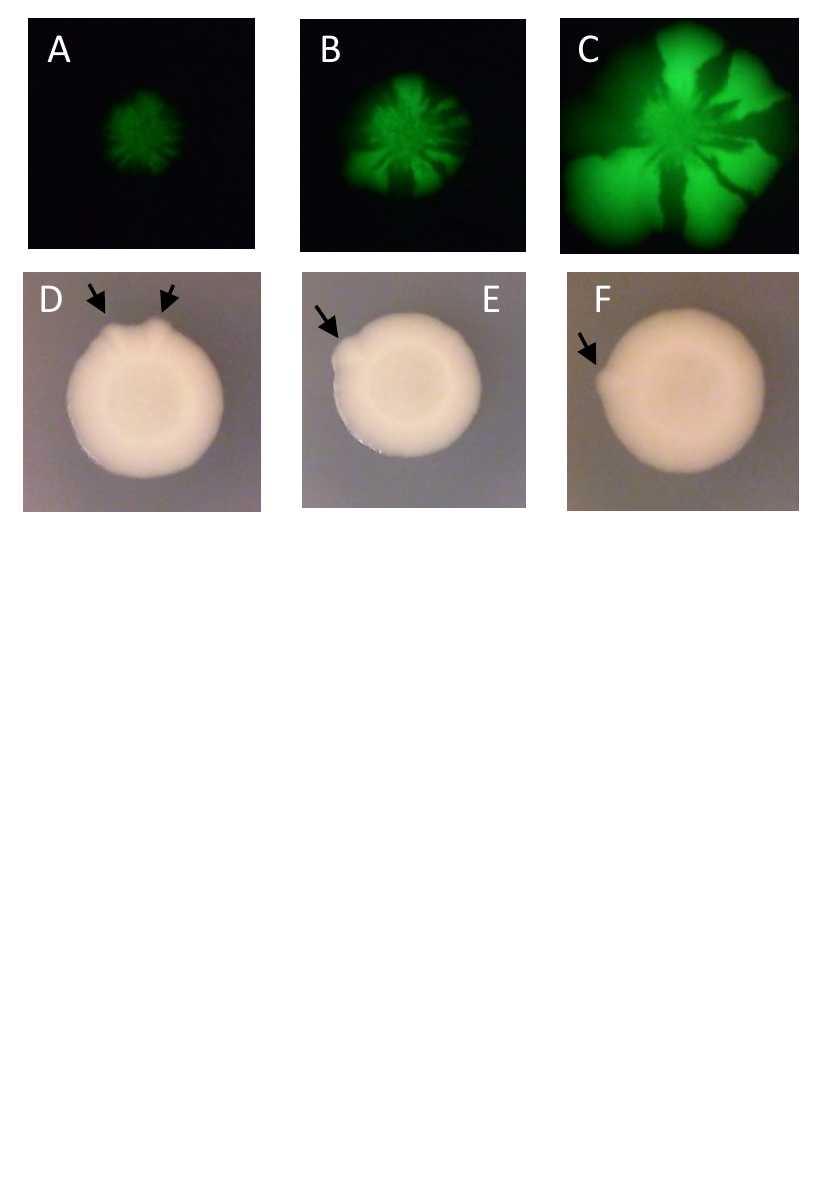}
	\caption{\label{fig:natural_sectors}A-C: sectoring patterns in a mixed colony of fluorescent and non-fluorescent {\it E. coli}. Images show the same colony at different times: $t_A=15,t_B=23,t_C=56$h. D-F: three examples of colonies (WT on LB+agar) which developed visible protrusions (``sectors'') of cells that outgrew the parent colony. The colony started from an initial ``coffee ring'' of bacteria deposited on agar (Fig. \ref{fig:static_cols}) and was incubated at 37$^{\circ}$C for 24h.}
\end{figure}

In the last 10 years, microbial colonies have become a popular model system to study the role of spatial structure on the dynamics of a population of organisms that expands in space into a new territory (``range expansion''  \cite{kirkpatrick_evolution_1997}). In a typical experiment, a mixture of two types of cells (``wild-type'' and ``mutant'') is deposited in a small droplet of liquid on the surface of an agar-filled Petri dish \cite{hallatschek2007genetic,hallatschek2010life,korolev2010genetic,korolev2012selective, mitri_resource_2016}. In the simplest case, the wild-type and the mutant have the same growth rate (fitness) but are genetically modified to express different fluorescent proteins which makes them easy to distinguish when illuminated with light of appropriate wavelength. After incubation, a colony with a distinct pattern of ``sectors'' of both types of cells is obtained (Figure \ref{fig:natural_sectors}A-C). Similar patterns are obtained if the colony is initiated from a single cell whose progeny can spontaneously switch to the ``mutant'' phenotype with small probability \cite{fusco_excess_2016}.

The emergence of sectors is the result of demographic fluctuations in the number of cells at the colony's edge. The phenomemon is similar to fixation/extinction of mutants in fixed-size population models such as the Moran process \cite{moran_random_1958}. Fluctuations cause some sectors to grow at the expense of other sectors that collapse, and mutants that stay at the front of the expanding colony are more likely to attain high frequencies. This is called ``gene surfing'' \cite{hallatschek2008gene}, and a surfing probability $P_{\rm surf}$ (a ``soft'' equivalent of the fixation probability \cite{nowak_evolutionary_2006}) can be defined as the probability that a mutant cell forms a macroscopic sector that stays at the expanding frontier. 

The number of sectors and their size depend on the surfing probability of beneficial mutations, which in turn correlates with the mutant's selective advantage $s$ defined as the ratio of the growth rates minus one, $s=g_{\rm mutant}/g_{\rm wild-type}-1$. If the mutant grows at the same rate as the wild type ($s=0$), it forms sectors of an angular size that does not increase in time once they have grown big enough that demographic fluctuations can be neglected \cite{hallatschek2010life}. For this reason, the number of sectors is finite in large colonies. However, if the mutant has a selective advantage $s>0$, the sectors become funnel-like; their angular size increases over time until eventually they completely occlude the wild type \cite{hallatschek2010life, korolev2012selective, antal_spatial_2015}.

The relationship between fixation probability and selective advantage in idealized models of biological evolution such as the Moran process is a classical result of population genetics \cite{nowak_evolutionary_2006}. In the well-mixed case, the fixation probability $P_{\rm fix}\cong s$ for $s\ll 1$. The proportionality $P_{\rm fix}\propto s$ generalizes to other models, modulo a numerical factor \cite{ewens_mathematical_2004,nowak_evolutionary_2006}. However, the equivalent relationship between $P_{\rm surf}$ and $s$ in real bacterial colonies is not well understood and is not universal. Models of expanding colonies predict that $P_{\rm surf}$ does not have to be linear in $s$ \cite{hallatschek2007genetic,hallatschek2010life}, and that it depends on the thickness and roughness of the growing layer of cells at the colony's edge \cite{gralka_allele_2016, farrell_mechanical_2017}. Computer simulations and experiments suggest that these quantities in turn are strongly affected by nutrient abundance \cite{bonachela_universality_2011,farrell_mechanically_2013}, cell shape \cite{gralka_allele_2016,smith_cell_2016}, and interactions between cells and the agar \cite{farrell_mechanical_2017}.

Experimental work described above uses genetically labelled cells to visualize the emergence of sectors. However, sectors of apparently faster-growing cells arise spontaneously in wild-type populations. The most popular microbe used in these experiments is the bacterium {\it E. coli} -- the  ``workhorse" of the microbiology lab. The bacterium is spherocylindrical, $\sim 0.8-1\mu$m in diameter and $\sim 2-4\mu$m in length \cite{neidhardt_book,growth_bacteria_book}. When cultured on agar plates with abundant nutrients, {\it E. coli} often produces a few sectors (Figure \ref{fig:natural_sectors}D-F) \cite{shapiro_significances_1995}. The surface of the sectors has a slightly different visual appearance to the rest of the colony, and the sectors protrude outward from the colony. We note that the bacteria are not motile in these experiments, so the presence of the sectors cannot be explained by switching to a motile phenotype. 

The stochastic nature of the sectors, their relatively low frequency, and the fact that they occur even when the colony grows from a single cell suggests that the process responsible for the sectors (i) is a genetic mutation or a rare epigenetic alteration (phenotype switch) that happens during colony expansion, and (ii) it enables cells to spread faster on agar plates and outcompete the parent strain. The selective advantage is unlikely to be caused by an increase in the exponential-phase growth rate (maximum growth rate when nutrients are abundant), because {\it E. coli} cannot double any faster on rich media than it already does (doubling time 20 min). It could however be that cells switch to a different, heritable metabolic state which makes them less sensitive to the decrease in nutrient concentration or the accumulation of waste products.

An alternative explanation is that the selective advantage of cells in such spontaneously produced sectors is conferred not by the difference in growth rates but by their ability to spread on agar. Since bacteria are non-motile, this change in spreading must be attributed to reduction in adhesion which would enable the cells to slide faster on the surface of agarose when being pushed by other growing cells. The idea that physical interactions between bacterial cells, and between the cells and the surface on which they grow, affect biological evolution, is relatively new to evolutionary biology but not surprising to a physicist. In particular, in our previous work we have used computer models to show that mechanical (physical) interactions significantly affect the surfing probability \cite{farrell_mechanical_2017} and horizontal gene transfer \cite{pastuszak_physics-explicit_2017} in bacterial colonies. 

Here we present experimental evidence that cell-surface interactions play a tangible role in population dynamics of {\it E. coli} colonies on agar plates.
Specifically, we perform competition experiments between a wild-type bacterium and a ``shaved mutant'' which adheres less strongly to agar. We show that although the growth rate of the mutant strain is the same as the non-mutated, wild-type strain in liquid cultures, the shaved strain forms colonies that expand faster on agar plates. When both strains grow next to each other, the shaved strain outcompetes the wild type. We hypothesise that this is because the mutant adheres less to the agar and is able to slide past wild-type cells and gain better access to nutrients.

\section{Results}
We created the shaved mutant AD32 by knocking out (deleting) two genes of the wild-type strain MG1655 of {\it E. coli}: fimA and fliF (Experimental Methods). FimA is the major subunit of the type-1 fimbriae (pili) \cite{barocchi_bacterial_2013}. Other components of the pilus are the adaptor proteins FimG and FimF, and the adhesin FimH attached at its distal end. While we have not deleted those genes, functional pili cannot be produced without fimA. FliF is a protein required for the formation of the MS ring which anchors the flagellum (an appendage used for swimming) in the cytoplasmic membrane \cite{berg_rotary_2003}. Without FliF, functional flagella cannot form; we confirmed by imaging dilute suspensions of AD32 in LB that these bacteria do not swim. We also created a  fluorescent strain RJA002 (a derivative of MG1655) with intact fimA and fliF genes. In what follows we shall refer to AD32 as the {\it shaved} strain, whereas RJA002 will be referred to as the {\it wild-type} strain.

\subsection{Shaved strain adheres less strongly to agar}
\begin{figure}
	\includegraphics[width=\columnwidth, trim=0 180 0 0]{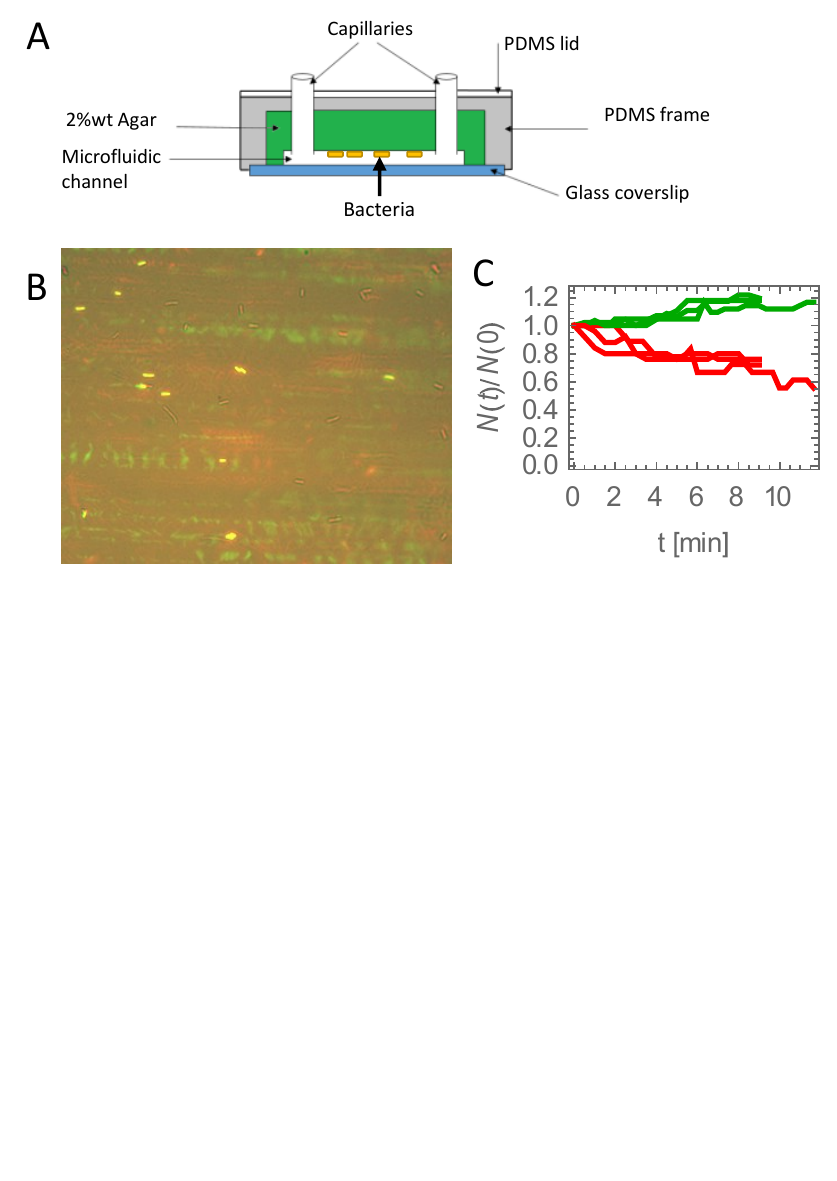}
	\caption{\label{fig:adhesion}Shaved cells adhere less to agar and are flushed away by flow. A: schematic of the experiment. B: image of a small region of the agar channel during flow (average over 15 images in a 20-second interval). Adhered bacteria (green = shaved AD104, red = wild type) are visible. ``Smears'' in the background come from moving cells which are being dragged by the flow. C: Number of cells $N(t)$ after $t$ minutes of flow relative to the initial number $N(0)$ for both types (notice the change of colour convention: green = wild-type, red = AD104).}
\end{figure}

We first sought to establish how well the shaved strain adheres to agar. We created an agar microfluidic channel (Fig. \ref{fig:adhesion}A) and filled it with a dilute 1:1 mixture of shaved and wild-type bacteria.
We used AD104 -- a brightly fluorescent version of AD32 (Methods) -- as the shaved strain in these experiments because it can be distinguished from the wild type more easily under the microscope.

Shortly after filling the channel with bacterial suspension, some bacteria attached to the agar (Fig. \ref{fig:adhesion}B). After 5 mins we flushed the channel for 10 mins (fluid velocity $\approx 10$\um/s at the distance of $5$\um\ from the surface). We reasoned that drag and shear forces from the flow would affect shaved bacteria more than wild-type cells and cause them to detach. Indeed, we observed shaved cells to detach from the surface, whereas the number of wild-type cells actually increased (Fig. \ref{fig:adhesion}C).

\subsection{Shaved and wild-type strain have the same growth rate in liquid media}
\begin{figure}
	\includegraphics[width=0.9\columnwidth, trim=0 100 0 0]{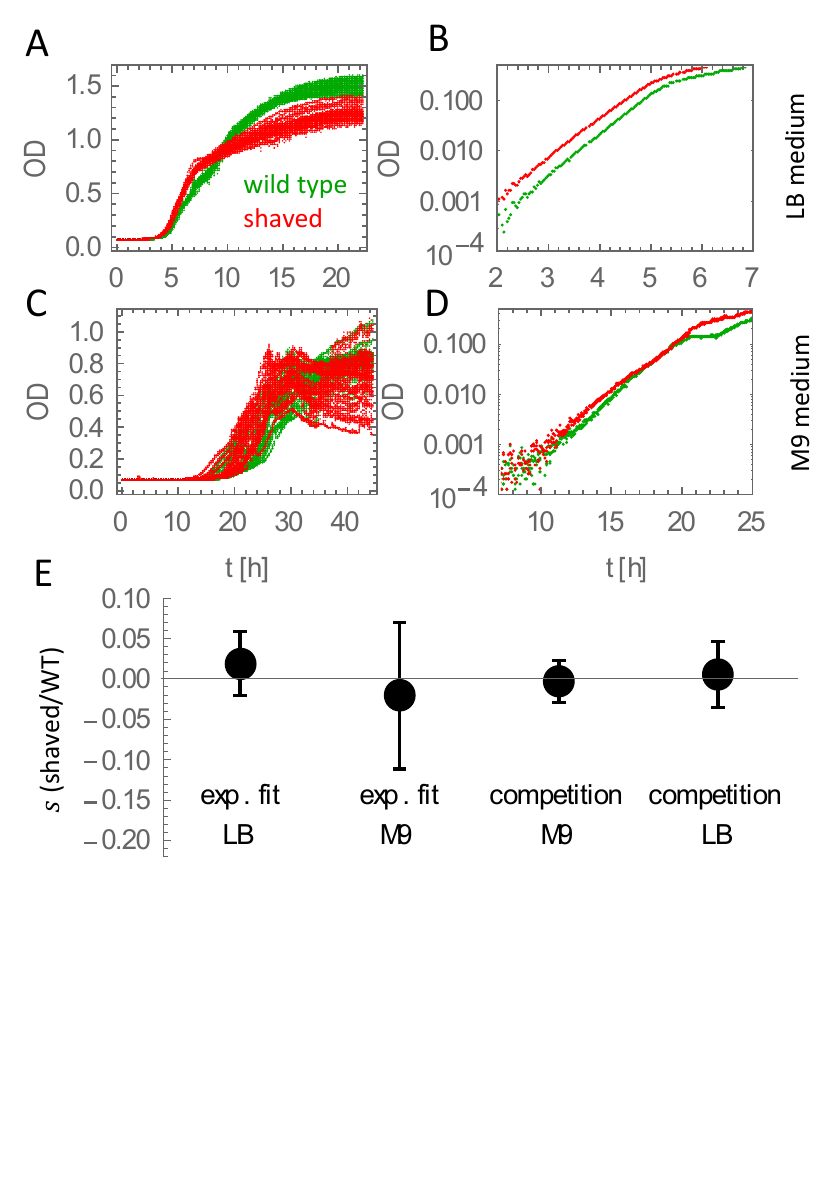}
	\caption{Wild-type and shaved bacteria grow similarly in liquid media. A: growth curves in LB (24 replicates for each strain). B: average growth curves in log-scale; exponential growth is clearly visible for OD $<0.1$. C-D: growth curves in M9. E: selective advantage $s$ of the shaved versus the wild-type strain obtained in two ways: Eq. (\ref{eq:expgrowth}) (``exp. fit'') and from the competition experiment. Error bars are standard errors (s.e.).\label{fig:s_in_liquid}}
\end{figure}

To confirm that genetic modifications to the shaved strain did not affect its growth rate in liquid media, we incubated both strains in a 96 well plate filled with growth media M9 or LB and recorded the optical density in each well over a 48h period. Figure \ref{fig:s_in_liquid}A,B shows LB growth curves. Variability between replicates is low, but curves for different strains differ slightly. This may be due to both strains clumping or adhering to walls of the plate in different ways. M9 growth curves display much higher variability (Fig. \ref{fig:s_in_liquid}C,D). We then extracted the exponential growth rate $g$,
\bq
	{\rm OD}(t)={\rm OD}(0)e^{gt},  \label{eq:expgrowth}
\eq
by fitting Eq. (\ref{eq:expgrowth}) to ${\rm OD}(t)$ for individual curves in their exponential phase (OD$<0.1$), and calculated the selective advantage $s=g_{shaved}/g_{WT}-1$ of the shaved strain over the WT strain. Figure \ref{fig:s_in_liquid}E shows that $s$ is close to zero for both growth media M9 and LB.

To check whether both strains continue to grow at the same rate at higher densities, and whether the differences between the curves from Fig. \ref{fig:s_in_liquid}A,C are artefacts specific to growth in a micro-plate, we performed competition experiments. We incubated a mixture of both strains in a well-aerated and vigorously shaken flask to saturation, and measured the initial and final frequency of cells of each strain. We then used Bayesian inference to estimate the distribution of selective advantage $s$ consistent with the observed frequencies.
Figure \ref{fig:s_in_liquid}E shows that the average $s$ is close to zero (within error bars) 
which agrees with no growth advantage being found from direct measurements of the growth rates.

\subsection{Shaved strain creates larger colonies on agar plates}
\begin{figure}
	\includegraphics[width=0.9\columnwidth, trim=0 55 0 0]{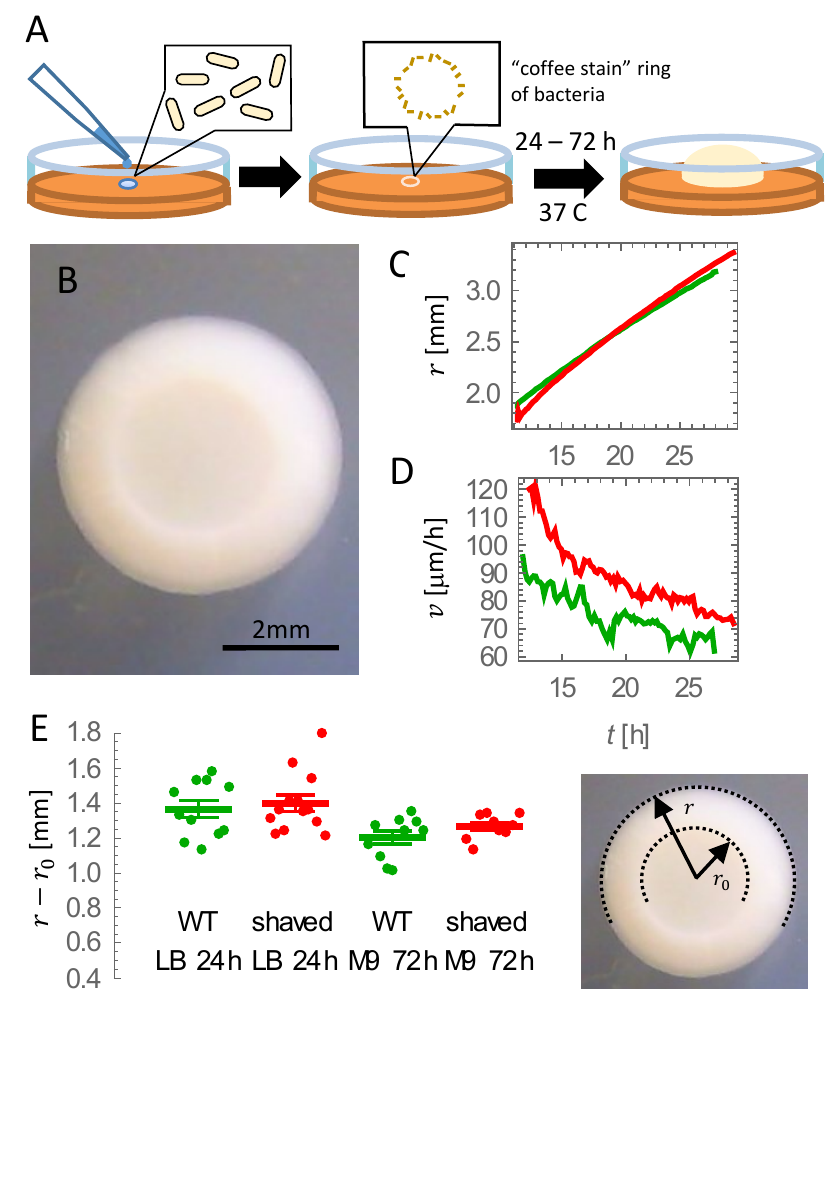}
	\caption{Growth of isolated colonies of the wild type and shaved bacteria on agar plates. A: schematic of the experiment. B: image of a typical colony (WT on M9 after 72h of incubation). C: radius $r$ of the colony as  a function of time $t$ from inoculation, for two continuously-imaged colonies: wild-type (red) and shaved (green). D: radial expansion speed $v$ as a function of time for the same two colonies. E: expansion distance $r-r_0$ where $r_0$ is the radius of the initial ring of cells. Error bars are s.e. \label{fig:static_cols}}
\end{figure}

We next grew isolated colonies of the shaved and wild-type strains on agar plates (Fig. \ref{fig:static_cols}A). Most colonies were round in appearance (Fig. \ref{fig:static_cols}B), with a distinct ring corresponding to the initial ``coffee-stain'' ring of deposited cells. We measured the speed of radial expansion of a few colonies by imaging them at regular intervals. Figure \ref{fig:static_cols}C shows the colony radius as a function of time, for two colonies of the shaved and wild-type strain. The shaved-strain colony expands faster (Fig. \ref{fig:static_cols}D) than the wild-type colony despite having a smaller initial radius.

We next determined the difference between the final and initial radii of colonies for both strains and the two growth media, M9 and LB, after incubating colonies for 24h (LB) or 72h (M9). Figure \ref{fig:static_cols}E shows a small difference in favour of the shaved strain; the distance travelled by the colony's front appears to be slightly larger than for the wild type. The spread in the data is however too big to draw a definite conclusion from this experiment.

\subsection{Shaved strain exhibits a selective advantage on agar}
\begin{figure}
	\includegraphics[width=0.9\columnwidth, trim=0 125 0 0]{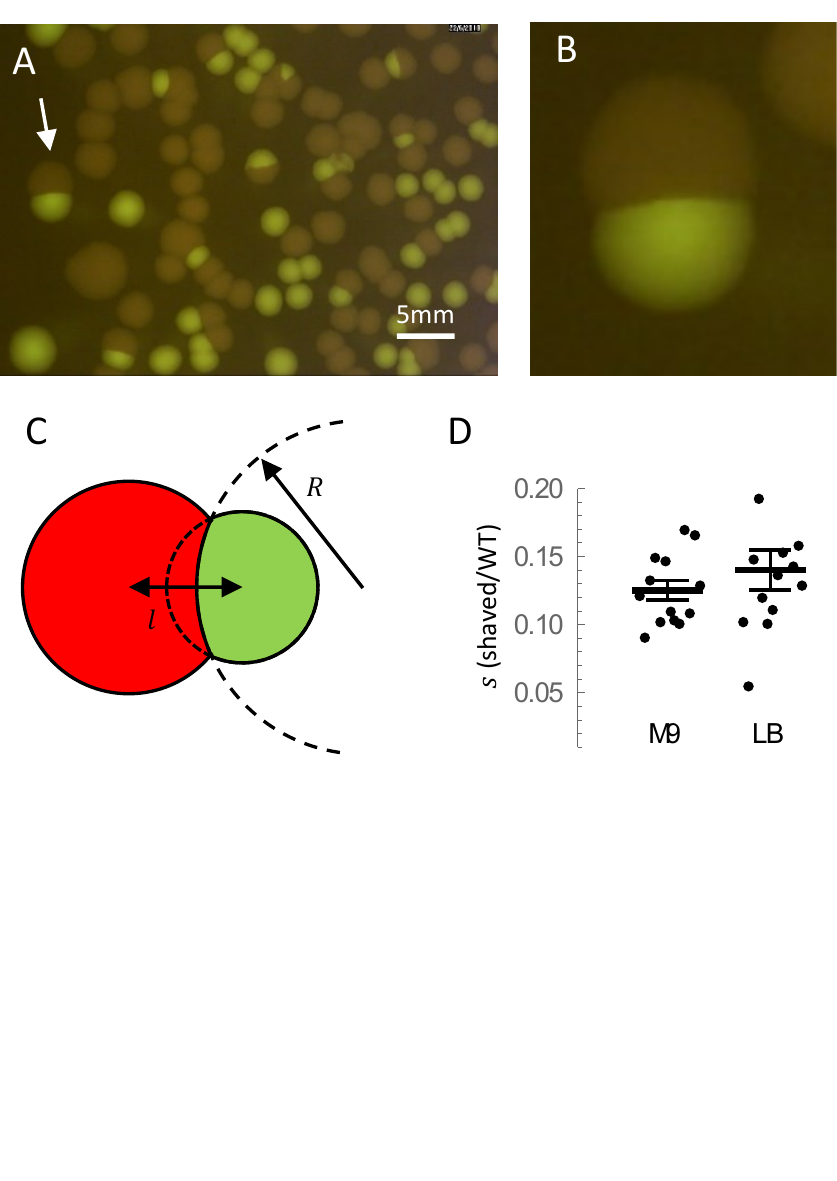}
	\caption{Colliding colonies assay. A: fragment of a Petri dish with colonies of the wild-type (brown) and shaved (green) bacteria. B: a close-up on a pair of colonies that collided. C: Definition of the radius of curvature $R$ of the collision interface. $l$ is the distance between the centres of the colonies. D: selective advantage $s$ estimated using Eq. (\ref{eq:sfromcoll}) for M9 and LB agar plates. Error bars are s.e. \label{fig:colliding}}
\end{figure}

To confirm that the difference in radial expansion velocities was due to different properties of the two strains and not e.g. differences in preparing agar plates or the growth medium, we inoculated agar plates with a diluted mixture (approx. 1:1) of cells of both strains. The high density of cells led to many collisions between colonies of different strains (Fig. \ref{fig:colliding}A). If both strains grew at the same rate, the interface between the two colliding colonies should be a straight line \cite{korolev2012selective}. We observed curved interfaces (Fig. \ref{fig:colliding}B) which means that the shaved strain expanded slightly faster than the wild-type strain. Assuming constant expansion velocity ($v_{WT}$ and $v_{S}$ for the wild-type and shaved strain, respectively), Eq. (18) from Ref. \cite{korolev2012selective} relates the radius of curvature $R$ of the interface to the initial distance $l$ of the two cells that initiated the colonies and the selective advantage $s=v_{S}/v_{WT}-1$ of the faster-expanding strain: $R = l \frac{1+s}{s(2+s)}$ (Fig. \ref{fig:colliding}C). This can be used to calculate $s$:
\bq
	s = \frac{\sqrt{l^2+4 R^2}+l-2 R}{2 R}. \label{eq:sfromcoll}
\eq
When we applied Eq. (\ref{eq:sfromcoll}) to our colliding colonies, we obtained $s=0.125\pm 0.007$ on M9 and $s=0.140\pm 0.015$ on LB (mean $\pm$ s.e.), i.e., the shaved strain has a statistically-significant advantage over the wild-type. 

\subsection{Computer simulations show that lower cell-surface friction leads to selective advantage}
The origin of the selective advantage cannot be explained by differences in the growth rate, but rather by different colony expansion velocities. This in turn can be attributed to differences in the interactions with agar for the two strains. To support this hypothesis, we performed computer simulations of growing colonies. We used the model from Ref. \cite{gralka_allele_2016}. Bacteria were represented as two-dimensional rods with spherical caps. Cells consumed nutrients diffusing in two-dimensional agar underneath the colony, elongated, replicated by binary fission, and repelled each other mechanically via contact forces (Hertzian model). 

\begin{figure}
	\includegraphics[width=\columnwidth, trim=0 255 0 0]{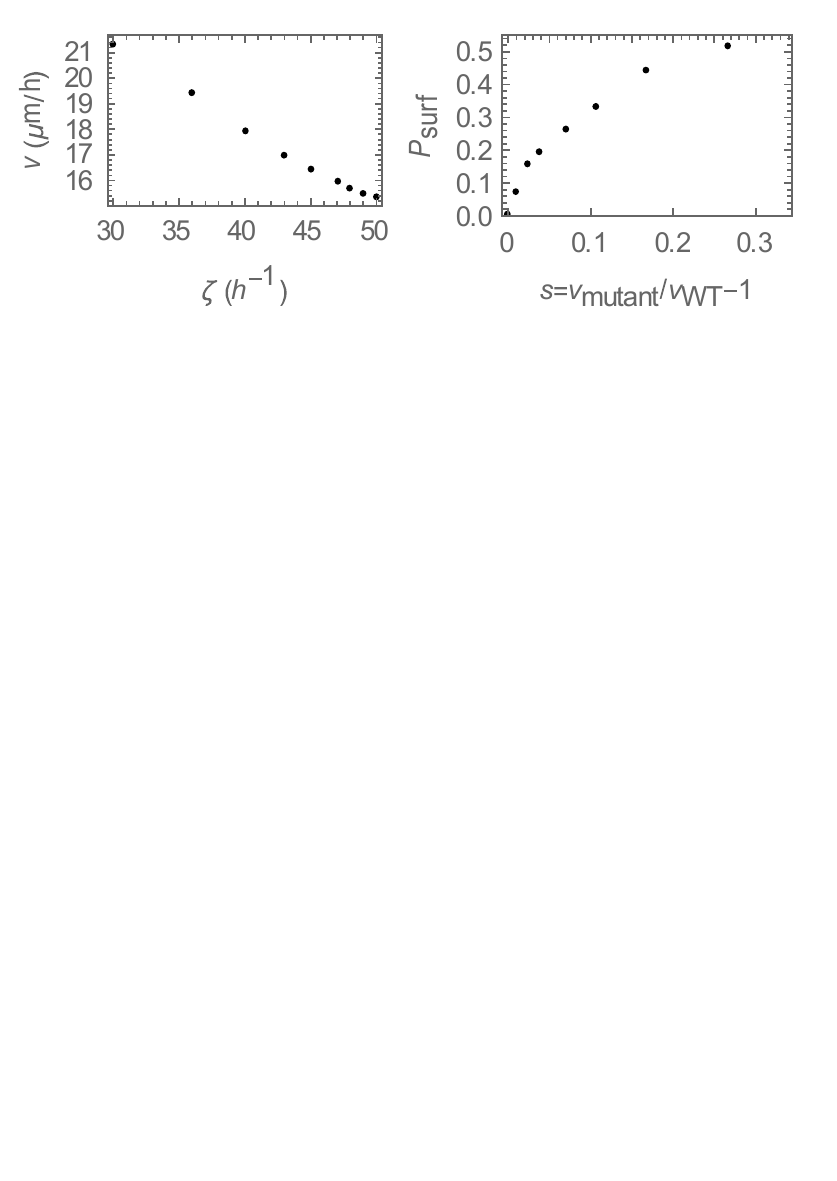}
	\caption{\label{fig:simulations}Computer simulations. A: front velocity as a function of the friction coefficient $\zeta$. The maximum $\zeta$ has been chosen to reproduce the average expansion velocity $v_{WT}=15$\um/h of our WT colonies growing on M9. B: Surfing probability $P_{\rm surf}$ versus $s=v_{mutant}/v_{WT}-1$.}
\end{figure}

Interaction with the agar surface was simulated as a drag force proportional to the velocity of the cell, $F=-\zeta l v$, where $l$ is the cell length. We varied the proportionality coefficient $\zeta$ to model different levels of friction caused by the presence/absence of fimbriae and flagella. All parameters were as in Ref. \cite{gralka_allele_2016}, except for the growth and uptake rates which we decreased three times to account for the reduced growth rate on M9. To save on the computation time, we only simulated a strip of cells of width $L=320$\um\ as in Ref. \cite{farrell_mechanical_2017}. To measure the speed of the colony's front, simulations were initiated from a line of cells and were run until the speed of the front stabilized. Figure \ref{fig:simulations}A shows that the front speed decreases with increasing friction coefficient $\zeta$. We can read off from this plot that a $13\%$ increase in the speed of the shaved mutant could  be explained by a $20\%$ reduction of the friction coefficient. Our model is of course highly idealised, so this value should be treated with caution. It shows, however, that a moderate decrease in friction is enough to produce a measurable difference in the expansion velocity.

To calculate the surfing probability $P_{\rm surf}$, we inserted mutants at random locations in the first line of cells from the steady-state configurations obtained in the speed measurement runs. We run the simulation until the growing layer was made only of one type (wild-type or mutant). The simulation was repeated 1000 times for each $\zeta$ and $P_{\rm surf}$ was estimated as the proportion of runs in which the mutant fixed in the growing layer. Figure \ref{fig:simulations}B shows that even a relatively small increase in the expansion speed of the mutant leads to $P_{\rm surf}\gg 0$.

\subsection{Shaved strain outcompetes the wild-type strain in mixed colonies}
\begin{figure}
	\includegraphics[width=0.9\columnwidth, trim=0 150 0 0]{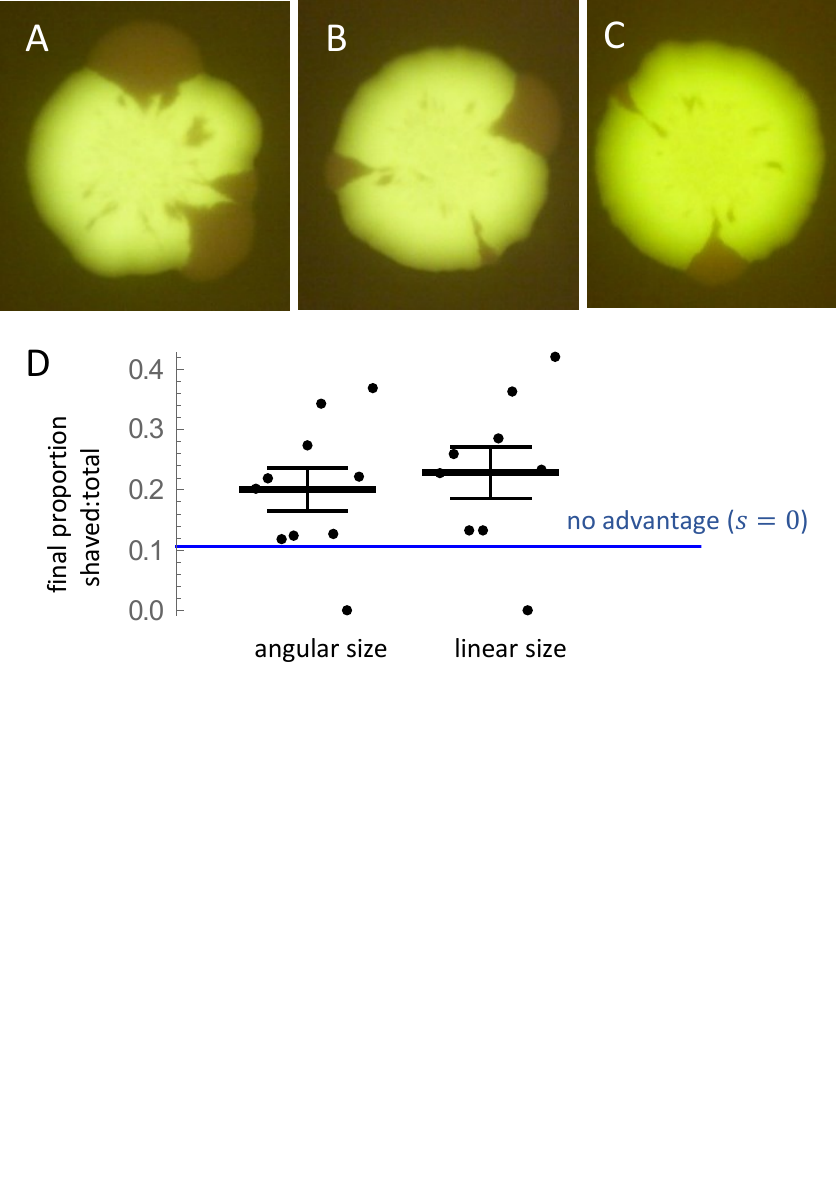}
	\caption{\label{fig:sectors}Sectors of the shaved strain in colonies of the wild-type strain. A-C: example colonies. The shaved strain forms funnel-like sectors ending with a buldge that protrudes from the colony. D: final proportion of the shaved strain estimated from the total size of all sectors of the shaved type. Errors are s.e. Blue line shows the expected proportion (equal to the initial proportion) in the absence of selection. }
\end{figure}

To confirm the results of computer simulations, we performed standing-variation experiments such as those in Fig. \ref{fig:natural_sectors}A-C. Figure \ref{fig:sectors}A-C shows images of colonies obtained from mixtures of the shaved and wild-type bacteria in proportion $1:8.4$. As expected, the shaved strain forms expanding sectors on agar plates. To estimate the final fraction of shaved cells at the colony's boundary, we used two methods: we either summed up angular or linear sizes of all sectors for each colony. In the absence of selection, the final fraction should remain close to the initial fraction of $0.106$. Figure \ref{fig:sectors}D shows that the fraction is higher than $0.2$, confirming that the shaved strain outcompetes the wild type.

\section{Conclusion and outlook}
We have shown that the lack of fimbriae and flagella enables {\it E. coli} to spread faster on agar, which confers a selective advantage in competition experiments. Faster spreading is likely to be caused by reduced adhesion to the agar, which in turn decreases friction between bacteria and the agar. 

Our results underscore the role of mechanical interactions in bacterial colonies from the population dynamics perspective and show that such interactions may be important for biological evolution of bacteria. Although the selective advantage of our shaved mutant is only $s\sim 13$\%, this is enough to significantly alter the frequency of the mutant in just one day (Fig. \ref{fig:sectors}).
It would be interesting to see if complete removal of all other surface adhesion proteins (surface antigens, curli, type IV pili, LPS) could lead to larger $s$ and even stronger selection.

{\it E. coli} is known to spontaneously switch between phenotypes expressing different levels of fimbriae and other surface appendages; this affects colony morphology \cite{hasman_antigen_2000}. Such rare, heritable variations in adhesion could be responsible for the emergence of sectors from Fig. \ref{fig:natural_sectors}. We have not directly tested this hypothesis, but our results support it.

Mutants that form bigger colonies on agar plates have been observed in {\it P. fluorescens} \cite{spiers_wrinkly-spreader_2007}, but they grow slower, and are eventually selected against. In contrast, we do not see any fitness cost in our shaved strain. Our results might thus suggest that reduced adhesion should be selected for. However, a Petri dish is a very different environment to what bacteria experience in nature, and one must not draw far-reaching conclusions from such simple experiments.

It therefore remains to be seen whether changes in adhesion and friction affect biological evolution in naturally occuring bacterial conglomerates such as biofilms \cite{donlan_biofilms_2002}. Biofilms grow on solid surfaces (the inside of a water tank, catheter, tooth enamel) immersed in a liquid (water, urine, saliva). Biofilms are of paramount interest to medicine and industry, and mechanical interactions play a very important role in the formation and growth of biofilms \cite{persat_mechanical_2015}. For example, adhesion mediated by type-1 pili enables {\it E. coli} to attach to uroepithelial cells in the urinary tract, invade them, and form intracellular colonies \cite{barocchi_bacterial_2013} which contribute to the persistence of urinary infections. Comparative genomics shows evidence of selection towards increased adhesion \cite{sokurenko_selection_2004}. However, it may be that intermediate levels of adhesion are actually better than no or too much adhesion. We believe that elucidating the coupling between mechanical interactions and biological evolution in biofilms will be an exciting area of future research.

{\it Acknowledgements.} 
BW was supported by a Royal Society of Edinburgh/Scottish Government Personal Research Fellowship. PH was supported by an EPSRC doctoral studentship. The work has made use of resources provided by the Edinburgh Compute and Data Facility (ECDF; www.ecdf.ed.ac.uk).

\section{Experimental methods}
\subsection{Bacterial strains}
We used three different strains of the bacterium {\it E. coli}. Yellow fluorescent protein (YFP) reporter strain RJA002 was made by P1 transduction from strain MRR of Ref. \cite{elowitz_stochastic_2002} into our laboratory stock of MG1655. YFP is expressed constitutively from the bacteriophage lambda PR promoter \cite{elowitz_stochastic_2002}.

Non-fluorescent ``shaved'' strain AD32 was constructed by sequential P1 transductions from JW1922 (BW25113 fliF from the Keio collection) followed by  JW4277 (BW25113 fimA from the Keio collection) into MG1655 \cite{baba_construction_2006}. The kanamycin resistance cassette was removed using  Flp recombinase  expressed in pCP20. Each step in strain construction was confirmed by PCR using a combination of kanamycin specific primers and gene specific primers. Fluorescent strain AD104 was obtained from AD32 by transforming it with pWR21, a plasmid constitutive for eGFP expression  \cite{pilizota_plasmolysis_2013}.

\subsection{Growth media}
M9 liquid medium with 0.2\% glucose was prepared by mixing 250ml of 4xM9 salts (28g Na2HPO4, 12g KH2PO4, 2g NaCL, 4g NH4Cl in 1L ultrapure water), 2ml of 1M MgSO4, 0.1 ml 1M CaCl2, and 10ml 20\% glucose, and adding ultrapure water to 1L. To create M9 in agar, 4xM9 salts and 2\% w/w agar (melted in a microwave) were put into a 60C water bath. After thermal equilibration, both solutions were mixed to yield M9 + 1.5\% agarose, and glucose was added to 20\% w/w. The medium was poured into Petri dishes to a depth of approx. 5mm. LB liquid medium was prepared according to Miller's formulation (10g tryptone, 5g yeast extract, 10g NaCl per litre). pH was adjusted to 7.2 with NaOH before autoclaving at 121C. To create LB in 1.5\% agar, agar (Oxoid, Agar  
Bacteriological, No. 1) was added before autoclaving.

\subsection{Growth rate in liquid media}
Growth curves.
A 96-well plate was filled with M9 or LB medium (200\ul\ per well) and inoculated with 5\ul\ of $1:10^3$ diluted exponential culture (OD$\approx 0.1$) of RJA002 or AD32.
The plate was incubated in a CLARIOstar plate reader (BMG Labtech) at 37$^{\circ}$C and optical density was measured every 2 mins. 

Competition experiment. Three 200ml flasks with 10ml M9+glucose or LB were inoculated with 10\ul\ of each strain (RJA002 and AD32) in the exponential-growth phase (OD $\approx 0.1$). We determined the frequency of each strain at $t=0$ and $t=24$h (LB) or $t=48$h (M9) (dense cultures, OD $>1.5$) by plating appropriately diluted samples on LB agar plates and counting fluorescent/non-fluorescent colonies (three replicates per flask). We used Bayesian inference to estimate the posterior probability distribution of the number of cells of each type in the initial ($N_{i,WT},N_{i,S}$) and final ($N_{f,WT},N_{f,S}$) samples. We then sampled $N_{i,WT},N_{i,S},N_{f,WT},N_{f,S}$ from the posterior distributions, and calculated $s=\log_2 [(N_{f,S}/N_{f,WT})/(N_{i,S}/N_{i,WT})]/G$ where $G$ was the number of generations, $G=\log_2 [(N_{f,S}+N_{f,WT})/(N_{i,S}+N_{i,WT})]$. The obtained distribution $P(s)$ was used to calculate the mean and standard error of $s$.

\subsection{Expansion velocities}
RJA002 and AD32 were grown overnight on LB and diluted into a fresh LB (50\ul\ into 10 ml). After approx. 1h of incubation at 37$^{\circ}$C in a shaken incubator (OD $\approx 0.1$), 1\ul\ of the suspension (approx. 10,000 cells/\ul\ as obtained by CFU count) was placed in the middle of an agar-filled Petri dish. The Petri dish was sealed with Parafilm to reduce evaporation, and continuously imaged in a 37$^{\circ}$C static incubator with a source of light (white LED) and a camera (Tecknet	C016 USB HD 720P Webcam) for up to 72h.

\subsection{Colliding colonies assay}
A $1:10^5$ dilution was made from the exponential-phase cultures of AD32 and RJA002 in phosphate-saline buffer (PBS). 500\ul\ of each strain was mixed together and 150\ul\ of the mixture spread onto a Petri dish. Cultures were incubated for 72h (M9) or 24h (LB). Each plate was photographed using a SafeImager blue light transilluminator and an amber filter. 

\subsection{Quantification of bacterial adhesion}
We created a 15mm x 2mm x 0.15mm agarose-lined micro-channel by pouring 2\% wt agarose + LB growth medium into a mold containing the negative of the channel. After excising a 4.7x2 cm area containing the channel and punching in- and outlet holes we fit it into a custom-made polydimethylsiloxane (PDMS) holder with a glass microscope slide at the bottom, and covered with a PDMS lid. Holes for glass capillaries were punched through the lid so that they lined up with the microfluidic channel. The edges and the capillaries were sealed using epoxy resin to create air-tight seal to prevent evaporation. 10\ul\ of overnight culture of AD104 and RJA002 was added to a conical flask containing 10mL LB, and incubated for 1h. Bacteria were injected into the device using a 1ml BD plastic syringe in a syringe pump (NE-1000, New Era Pump Systems Inc) connected to the device by an elastic tube. Cells were imaged using a Nikon epifluorescent microscope with a 20x dry objective in both brightfield and GFP channels. Cells were allowed to adhere for 5 min and then attempted to flush away with a 5$\mu$L/min flow. Images were recorded at $\approx 1$s intervals. 

\subsection{Sectoring experiments}
A 1:8.4 mix of AD32:RJA002 (proportion estimated from ODs of initial cultures) was prepared from the exponential-phase cultures. 1\ul\ of the mixture was placed in the centre of an LB-agar Petri dish as described in {\it Expansion velocities}. Plates were incubated at 37$^{\circ}$C for up to 48h and photographed using the transillumination box/amber filter. 

\subsection{Image analysis}
Images were analysed using ImageJ and Wolfram Mathematica. After tresholding, the JFilament2D plugin \cite{smith_segmentation_nodate} was used to trace the perimeter of each colony. A list of coordinates (700-1000) determining the perimeter was imported into a custom-written Mathematica script. Colony area was obtained by converting the points to a convex polygon. The radius was calculated as the average distance between the points and the centre of mass (CM) of the colony. 
The radius of curvature $R$ of the collision front was obtained by tracing points at the inteface using JFilamend2D \cite{ojkic_cell-wall_2016}; the region of smallest curvature was used to calculate $R$ using a Mathematica script. To find the distance between the colonies we manually fitted two circles to the colliding colonies and calculated the distance between their centres. 


\end{document}